\documentclass[12pt]{iopart}
\usepackage{epsf}
\begin{document}

\title {FDTD Simulation of Exposure of Biological Material to 
Electromagnetic Nanopulses}

\author{Neven Simicevic\dag \footnote[3]{To whom correspondence should be addressed Louisiana Tech University, 
PO Box 10348, Ruston, LA 71272, Tel: +1.318.257.3591, Fax: +1.318.257.4228, 
E-mail: neven@phys.latech.edu} \ and  Donald T. Haynie\ddag}

\address{\dag\ Center for Applied Physics Studies, Louisiana Tech University,
 Ruston, LA 71272, USA}

\address{\ddag\ Center for Applied Physics Studies, Biomedical Engineering 
 and Institute for Micromanufacturing, Louisiana Tech University,
 Ruston, LA 71272, USA}

\begin{abstract}

Ultra-wideband (UWB) electromagnetic pulses of nanosecond duration, or nanopulses,
are of considerable interest to the communications industry and are being explored 
for various applications in biotechnology and medicine.  The propagation of a 
nanopulse through biological matter has been computed in the time domain using the 
finite difference-time domain method (FDTD). The approach required existing Cole-Cole
model-based descriptions of dielectric properties of biological matter to be 
re-parametrized using the Debye model, but without loss of accuracy.  
The approach has been applied to several tissue types.  Results show that the 
electromagnetic field inside a biological tissue depends on incident 
pulse rise time and width.  Rise time dominates pulse behavior inside a 
tissue as conductivity increases.  It has also been found that the amount of 
energy deposited by 20 $kV/m$ nanopulses is insufficient to change the 
temperature of the exposed material for the pulse repetition rates 
of 1 $MHz$ or less.

\end{abstract}

\pacs{87.50.Rr, 87.17.–d, 77.22.Ch, 02.60.–x}


\maketitle

\section{Introduction}
A facility for bioelectromagnetics research has recently been established at 
Louisiana Tech University (LA Tech) through sponsorship by Air Force Office of 
Scientific Research. LA Tech leads a multi-university collaboration in this area 
which involves three other institutions in north Louisiana: Grambling State 
University, University of Louisiana at Monroe, and Louisiana State University-Health 
Sciences Center, Shreveport.  Current focus of research is 
bioeffects of non-ionizing ultra-wideband (UWB) electromagnetic (EM) pulses
of nanosecond duration, or nanopulses.  The research program encompasses experimental
studies of biological matter, equipment design and fabrication, 
and computational modeling.  Goals of the research include providing a sound basis 
for nanopulse exposure safety standards.

The literature on UWB radiation is extensive \cite{Tay95}.  In the present work, a
nanopulse is a rapid, transient change in amplitude, 
from a baseline to peak, followed by a relatively rapid return to baseline.  
It is a short duration, high-intensity burst of electromagnetic energy.
In the LA Tech bioelectromagnetics facility, fondly known as the Nanopulse Factory,
a typical nanopulse has a width of 1-10 $ns$, a rise time of 
$\sim$ 100 $ps$, and an amplitude of $\sim$ 20 $kV/m$.

Extensive research has been done on biological effects of EM fields.  
Detailed descriptions are provided in Reference \cite{Polk95}.  
Bioeffects of nanopulses, however, may be qualitatively different from those 
of narrow-band radiofrequencies. The LA Tech-led collaboration is currently testing 
nanopulse bioeffects using a range of model systems.  
At the cellular level this includes $E. \; coli$, photosynthetic bacteria, 
bovine red blood cells, bovine platelets, mouse hepatocytes, mouse mammary 
epithelial cells, and human dermal fibroblasts; that is, both prokaryotes and 
eukaryotes.  The main sub-cellular model is horseradish peroxidase.  
A whole animal model is $C. \; elegans$.

\begin{figure}
\begin{center}
\mbox{\epsfxsize=4.in\epsfbox{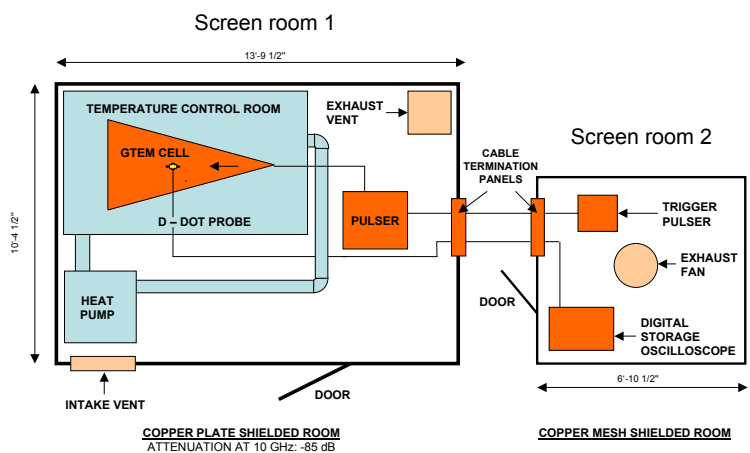}}
\caption{\label{fig01} Schematic of LA Tech nanopulse exposure facility. 
It consists of two Faraday cages, the GTEM cell and pulser in one and the 
measuring/recording instruments in the other.}
\end{center}
\end{figure}

The basic exposure equipment consists of a pulse generator, a parallel-plate 
transmission line ($e.g.$ gigahertz transverse electromagnetic mode or GTEM cell), 
measuring/recording instruments, and a radiofrequency enclosure 
(screen room, Faraday cage).  A schematic is shown in Figure~\ref{fig01}. 
Output of a commercial or home-built nanosecond pulse generator \cite{Sun04} 
is fed into the GTEM cell or a home-built parallel-plate capacitor, through 
which the pulse propagates virtually unperturbed to the position of the sample. 

Pulse generator output is measured and recorded using a digital 
storage oscilloscope. Nevertheless, it is a challenge to make accurate 
real-time measurements of the electric field in an exposure chamber in the 
vicinity of the sample, and it is practically impossible to measure the field 
inside the sample in real time. To find the field inside a 
sample, which is what one cares about, it is necessary to consider 
a computational approach.

The interaction of short EM pulses and biological matter 
has not been modeled in such detail as the interaction of radio frequency radiation.
A number of computational approaches exist for modeling the experimental apparatus, 
biological cell, and cellular environment, and the EM interaction mechanisms and 
their effects \cite{Polk95}. The complexity of any realistic situation requires 
a numerical rather than an analytical approach. The latter, however, 
should be taken in parallel with 
the former, since the dynamic range of the problem could span many orders of 
magnitude in some physical quantities and an ``external" check on computational 
method is needed.  In the case of a biological cell, for example, the length 
scale ranges over nine orders of magnitude, from the thickness of the 
plasma membrane to the size of the exposure chamber.  This represents a 
considerable challenge for any numerical method.

For the calculations described in the present work, finite-difference time domain 
FDTD was applied.  
This method of solving  Maxwell's equations is relatively simple, can easily 
deal with a broadband response, has almost no limit in the description of 
geometrical and dispersive properties of the material being simulated, is 
numerically robust, and is appropriate for the computer technology of today. 
Originally introduced by Kane Yee in the 1960s \cite{Yee66}, FDTD was developed 
extensively in the 1990s \cite{Sad92,Kunz93,Sull00,Taf00}, owing in part to the 
increasing availability of fast computers. In this paper we describe 
FDTD calculations of the EM field inside samples exposed to nanopulses in a GTEM cell.  
The EM properties of the environment are included in the calculation to the 
fullest extent. The object is to advance understanding 
of dominant mechanisms of interaction of nanopulses with biological structures.

\section{Computational Inputs}

\begin{figure}
\begin{center}
\mbox{\epsfxsize=4.in\epsfbox{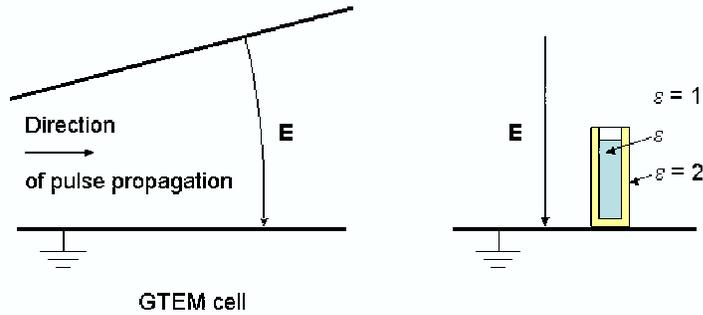}}
\vspace*{- 0.5cm}
\caption{\label{expo} Geometry of exposed sample: cuvette inside the GTEM cell.}
\end{center}
\end{figure}

In order to characterize the response of a biological system to an EM pulse, 
two important quantities must be known with a reasonable degree of precision: 
the value of the field surrounding the system and in the system, and the extent 
of conversion of EM energy into mechanical or thermal energy, both in the system 
itself and in the surroundings.  FDTD has been applied for this purpose, and an original 
set of computer programs has been developed at LA Tech to compute the 
EM field in any dimension for almost any choice of geometry and EM properties of 
a material. Some of the computations were performed using a 3-dimensional 
model, the results presented here, however, were obtained using 
2-dimensional FDTD. The approach was based on the following dimensions: 
samples in a cuvette (1 $cm$ $\times$ 1 $cm$ $\times$ 4.5 $cm$, with 1mm 
thick walls), and a GTEM cell in which 
exposure occurs (8 $cm$  $\times$ 8 $cm$  at inlet, 58 $cm$  $\times$ 58 $cm$ 
at absorbing cones, and 100 $cm$ long). 2-dimensional FDTD reduces the computation 
time without compromising essential features of the solution.  Geometry of the 
exposed sample is shown in Figure~\ref{expo}.

Each calculation depends on the shape of nanopulses fed into the GTEM cell,
defined geometrical properties of the exposed ``system", and its dispersive or 
dielectric properties (including conductivity).
It was important that each property be both realistic and appropriate for 
numerical simulation.  Further details of each feature are given in the 
following subsections.

\subsection{Electromagnetic Pulse Inside GTEM Cell}

The EM field of a nanopulse inside a GTEM cell can be measured 
when the cell is empty \cite{Bao97}.  
FDTD calculation of pulse propagation through a flared transmission line 
shows that the shape of the pulse is 
preserved as it propagates and, as expected, only the amplitude decreases.  
This agrees with the results of work done at Brooks Air Force Base 
(now Brooks City-Base) on modeling a GTEM cell \cite{Samn99}.
The pulse in a GTEM cell can be described as a 
double exponential function:

\begin{equation} E=E_{0}(e^{-\alpha t}-e^{-\beta t}),
\label{dubexp}
\end{equation}
where $E_{0}$ is pulse amplitude and $\alpha$ and $\beta$ coefficients 
describing pulse rise time, fall time, and width.  Parameters that describe pulse 
shape in the empty GTEM cell at LA Tech in the vicinity of the region under test 
(sample position) are $E_{0} = 18.5 \; kV/m$, $\alpha = 1. \times 10^{8} \; s^{-1}$, 
and $\beta = 2. \times 10^{10} \; s^{-1}$.  
This pulse, having a rise time of 150 $ps$ and width of 10 $ns$,
was the input in the present work.

\subsection{Geometrical Properties of Exposed Sample}

\begin{figure}
\begin{center}
\mbox{\epsfxsize=3.in\epsfbox{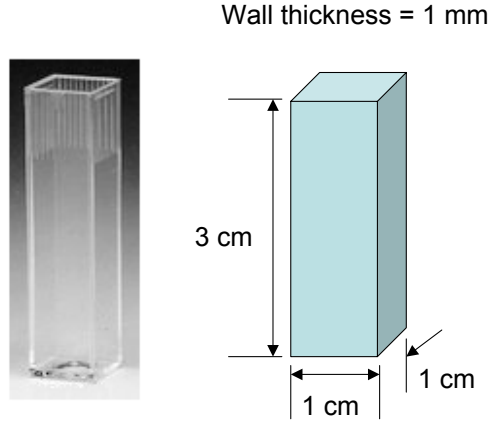}}
\caption{\label{fig05} Shape and dimension of a cuvette used in experiment and 
studies described here.}
\end{center}
\end{figure}

Most biological specimens in experiments in the LA Tech-led research program 
consist of mammalian cells or microorganisms (length $\leq 1 \; mm$).  
This size is small in comparison to the dimensions of the GTEM cell and will 
not perturb the general character of the EM field. In other words, the character 
of the field, anywhere in a GTEM cell except in the vicinity of the
sample, will be roughly the same as in an empty cell. The largest object in 
the GTEM cell during an experiment is the sample container. Ordinarily this will 
be a polystyrene cuvette, whose shape and dimensions are shown in 
Figure \ref{fig05}, a Petri dish, or a 96-, 48-, or 8-well plate.

Other considerations must be made when describing geometrical properties of 
an object in an FDTD simulation.  The method requires space and time to be 
discretized.  The discretization of space is done by means of Yee cells,
cuboids having edge lengths $\Delta x$, $\Delta y$, and $\Delta z$. 
If $\Delta x=\Delta y=\Delta z$, a Yee cell represents a discrete cube of space.  
The discretization of time is obtained from the size of the Yee cell by imposing 
the Courant stability criterion:

\begin{equation}
\Delta t \leq {1 \over {c \sqrt{(\Delta x)^{2}+(\Delta y)^{2}+(\Delta z)^{2}}}},
\end{equation}
where $c$ is the speed of light.

Yee cell must be small enough not to distort the shape of the sample container,
has to account for the full frequency range of the EM pulse, and must be 
large enough for the time step to be practical for overall computation.
Its size is related to the highest frequency which needs to be 
considered, $f_{max}$, by an accepted rule

\begin{equation}
\Delta x \simeq {c \over {10 \; f_{max}}},
\end{equation}
where $c$ is the speed of light and $f_{max}$ is a cut-off frequency above which 
the calculation becomes unreliable for the chosen cell size.  In the present work 
the maximum considered frequency was $f_{max}= 100 \; GHz$, which required 
the size of the Yee cube edge lengths to be 
$\Delta x=\Delta y=\Delta z \simeq 0.3 \; mm$.  
A cell edge length of 1/4 $mm$ satisfies the frequency criterion and is small 
enough to describe the shape of the sample, and derived time step satisfying 
the Courant stability criterion, $\Delta t \simeq 0.6 \; ps$, is large enough 
to allow the entire calculation to be performed in about 50,000 steps.

It is not always possible to achieve optimal agreement between geometrical and
physical descriptions of a situation. Fortuitous circumstances in the present 
work minimized the number of computational operations, eliminated need of 
additional approximations, and allowed the entire 2-dimensional FDTD calculation 
to be performed on a modern computer in about 10 minutes.

\subsection{Dielectric Properties of Exposed Sample}

Dielectric properties of the exposed sample were treated using a recursive 
convolution scheme \cite{Lueb90}.  Briefly, a relation between the 
electric flux density,
$\vec D$, and the electric field strength, $\vec E$, at points in the material 
at which the field was calculated, for a monochromatic EM wave, is

\begin{equation}
\vec D({\omega}) = \epsilon(\omega) \vec E({\omega}).
  \label{deE}
\end{equation}
Electric permittivity $\epsilon(\omega)$ is a function of frequency $\omega$
of the monochromatic wave. FDTD requires a connection between $\vec D$ and $\vec E$ 
in the time domain, which can be found by 
Fourier transformation of Equation~\ref{deE}.  The result 
can be written as \cite{Jackson99}

\begin{equation}
\vec D(t) = \epsilon_{0} \vec E(t)  
+ \epsilon_{0} \int_{0}^{t} \chi(\tau) \vec E(t-\tau)\;d\tau.
  \label{Dt}
\end{equation}
where $\epsilon_{0}$ is the permittivity of free space, and $\chi(\tau)$, 
the electric susceptibility of a material, is described by 
the following Fourier transform:

\begin{equation}
 \chi(\tau) = {1 \over {2\pi}} \int_{-\infty}^{+\infty} 
 ({\epsilon(\omega)/\epsilon_{0}}+1)  e^{-i\omega t} d\omega.
  \label{ksift}
\end{equation}
In FDTD all physical quantities are discretized and

\begin{equation}
\vec D(t) \mapsto \vec D(n\Delta t) 
= \epsilon_{\infty} \epsilon_{0} \vec E(n\Delta t)
+ \epsilon_{0} \int_{0}^{n\Delta t} \chi(\tau) \vec E(n\Delta t-\tau)\;d\tau.
  \label{Dtd}
\end{equation}
The quantity $\epsilon_{\infty}$ describes the property of the material at 
frequencies approaching infinity, and $n$ is a time step of length $\Delta t$.  
Without going into details of FDTD, which in any case can be found in 
References \cite{Lueb90,Lueb91,Lueb92}, the value of each vector component 
in Equation~\ref{Dtd} at time step $n$ can be written in discrete form as

\begin{equation}
D^{n} = \epsilon_{\infty} \epsilon_{0} E^{n}
+ \epsilon_{0}  \sum_{m=0}^{n-1} E^{n-m} \chi_{m},
  \label{Dtdd}
\end{equation}
where
\begin{equation}
\chi_{m}= \int_{m\Delta t}^{(m+1)\Delta t} \chi(\tau) \;d\tau.
\label{chi1}
\end{equation}

EM properties of a biological material are normally expressed in terms 
of frequency-dependent dielectric properties and conductivity.  
They have been measured and modeled for over a 100 years, and a great deal of 
information on them is available in the literature \cite{Polk95}.  Data used in the 
present work are from References \cite{Gab96} and \cite{GabGab96}, 
where the measured values of 45 tissues were 
parametrized using the Cole-Cole model:

\begin{equation}
\epsilon(\omega) = \epsilon_{\infty} + \sum_{k=1}^{4}
{\Delta \epsilon_{k} \over {1+(i\omega\tau_{k})^{1-\alpha}}} 
+ {\sigma  \over i\omega \epsilon_{0}},
\label{CC}
\end{equation}
where $i=\sqrt{-1}$.  Permittivity in the terahertz frequency range 
$\epsilon_{\infty}$, drop in permittivity in a specified frequency range 
$\Delta \epsilon_{k}$, coefficient $\alpha$, relaxation time $\tau$, 
and the ionic conductivity 
$\sigma$, constitute up to 14 real parameters of the fit. This approach can 
generally be used with confidence for frequencies above 1 $MHz$ \cite{Gab96}, 
the frequency range of interest in nanopulse bioeffects study.  
A plot of all the fit curves \cite{Gab96} reveals similarities of the 
dispersive properties of the various tissues.

\begin{figure}
\begin{center}
\mbox{\epsfxsize=3.in\epsfbox{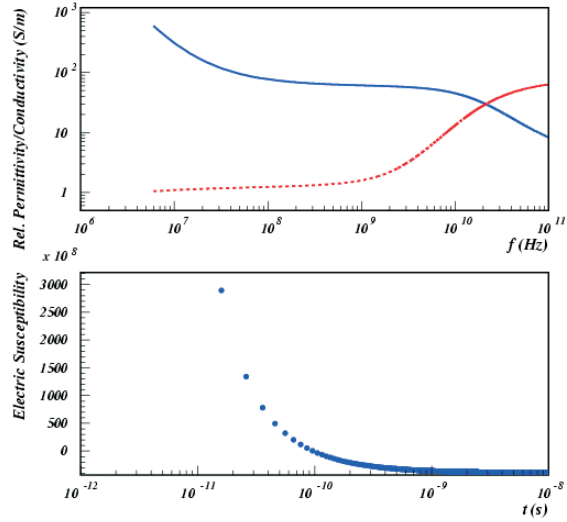}}
\vspace*{- 0.5cm}
\caption{\label{bldias} Top, relative permittivity (solid line) and 
conductivity (dashed line) of blood in the frequency range $\leq$ 100 $GHz$, 
calculated by Equation~\ref{CC} as parametrized in 
References \cite{Gab96} and \cite{GabGab96}.
Bottom, the electric susceptibility obtained by numerical Fourier transformation 
of the same equation.}
\end{center}
\end{figure}

While formally the electric susceptibility is just a Fourier transformation of 
Equation~\ref{CC}, the transformation is hardly easy \cite{Su04} and can only be 
achieved numerically. An example of a numerical Fourier transformation of a 
Cole-Cole expression, Equation~\ref{CC}, for blood
is shown in Figure~\ref{bldias}. Although this simple function can be 
modeled with just one free parameter, its application is problematic.

Cole-Cole parametrization can provide a useful empirical description of the 
dielectric properties of tissues over a broad frequency range.  
This model, however, does not reflect a specific underlying physical 
mechanism, as it is apparent from the divergence of $\epsilon(\omega)$ as the 
frequency goes to infinity when it should go to unity \cite{Land60}.  In addition, 
the components of the electric displacement $\vec D$, 
are calculated as a convolution of the electric field 
and material susceptibility, Equation~\ref{Dt}.  The response of a material 
to an external EM pulse is very fast.  Susceptibility, as shown in 
Figure~\ref{bldias}, is largest at the beginning of the response.  
Hence, precisely in the most 
important region for evaluating the integral in Equation~\ref{Dt} information on 
susceptibility will not exist.  The time step in calculating $\vec D$ using 
Equation~\ref{Dtdd} was 0.6 $ps$.  The first several steps of the computation 
therefore required the use of an extrapolated value of susceptibility.  
Because the Cole-Cole expression does not describe a physical mechanism, 
making such extrapolation has dubious validity and could represent a substantial 
source of error.  

There is another difficulty in applying the Cole-Cole 
parametrization.  Numerically, the electric displacement is calculated by 
Equation~\ref{Dtdd} as part of the overall Yee 
algorithm \cite{Yee66,Lueb91,Bui91,Lueb92}.  Evaluation of the integral 
in Equation~\ref{chi1} for all Yee cells at each time step, however, 
will be extremely time consuming for even the most effective integration techniques.

Both problems - extrapolation of susceptibility and numerical evaluation of 
Equation~\ref{chi1} - are more satisfactorily solved if Debye parametrization 
is substituted for Cole-Cole parametrization.  The Debye model describes 
relaxation of a material at the molecular level using an exponential function 
defined by a relaxation time $\tau$.  In place of Equation~\ref{CC}, 
neglecting conductivity $\sigma$ for the moment, dielectric properties of a 
material can be described as

\begin{equation}
\epsilon(\omega) = \epsilon_{\infty} + \sum_{k=1}^{N}
{ \Delta \epsilon_{k} \over {1+i\omega\tau_{k}}}=\epsilon_{\infty} +\sum_{k=1}^{N} \chi_{k}(\omega),
\label{Dpar}
\end{equation}
where $N$ is the number of independent first-order processes.  Response of 
the dielectric material to an external field in the time domain can be 
obtained by Fourier transformation of 
each independent first-order process $\chi_{k}(\omega)$ in Equation~\ref{Dpar}:

\begin{equation}
\chi_{k}(t)={ \Delta \epsilon_{k} \over \tau_{k}} \; e^{-t/\tau_{k}}, 
\; \;  {t \geq 0}.
\label{Dpart}
\end{equation}
where $\tau_{k}$ is the relaxation time for process $k$.

As to static conductivity $\sigma$, it is defined in the time domain as 
the constant of proportionality between the current density $\vec J$ and 
the applied electric field $\vec E$
as $\vec J = \sigma \vec E$. It is important to mention that its implementation 
in FDTD does not require additional or different Fourier transforms \cite{Kunz93}.  
The dependence of $\vec J$ on $\vec E$ in the conductive material is simply

\begin{equation}
\vec J = \sigma \vec E 
+\sum_{k=1}^{N} { {\Delta \epsilon_{k} \epsilon_{0}} \over \tau_{k}} 
\; e^{-t/\tau_{k}} \vec E, \; \;  {t \geq 0}.
\label{JE}
\end{equation}
The second term represents the effects of dielectric properties of the material.

The advantage of Debye parametrization becomes clear when evaluating 
Equations~\ref{Dtdd} and \ref{chi1}. After including the permittivity 
from Equation~\ref{Dtdd} in 
Equation~\ref{chi1}, it follows, for each independent first-order process, that

\begin{equation}
\chi_{m+1} 
= { \Delta \epsilon \over \tau} \int_{(m+1)\Delta t}^{(m+2)\Delta t} e^{-t/\tau}\; dt 
= \Delta \epsilon e^{-(m+1)\Delta t/\tau}(1-e^{-\Delta t/\tau})
= e^{-\Delta t/\tau} \chi_{m}.
  \label{chi2}
\end{equation}
From this it follows that the permittivity at time step $(m+1)$ is simply 
the permittivity at time step $m$ multiplied by a constant.  
A detailed description of this approach is given in Reference \cite{Kunz93}.

\begin{figure}
\begin{center}
\mbox{\epsfxsize=3.in\epsfbox{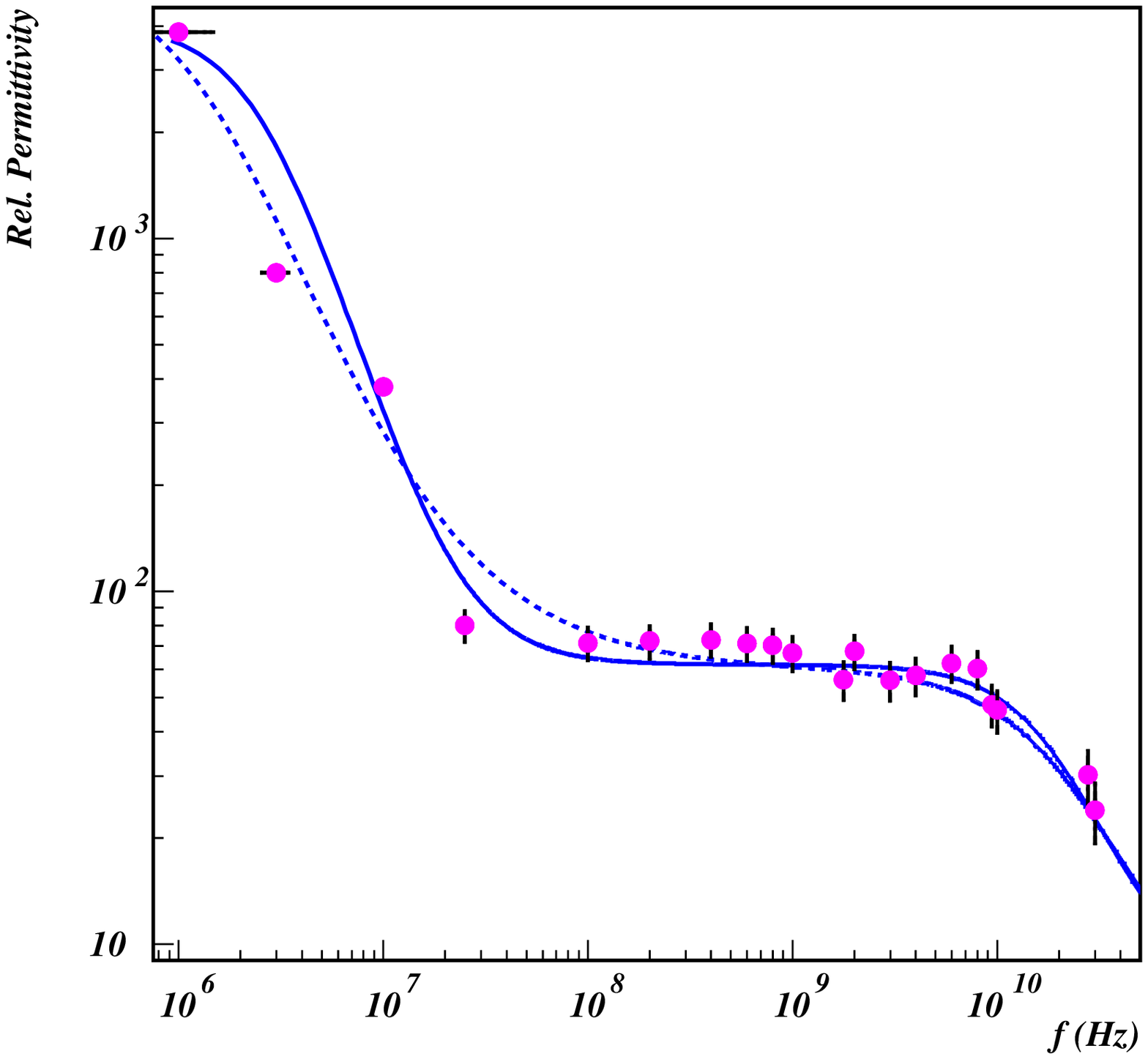}
\epsfxsize=3.in\epsfbox{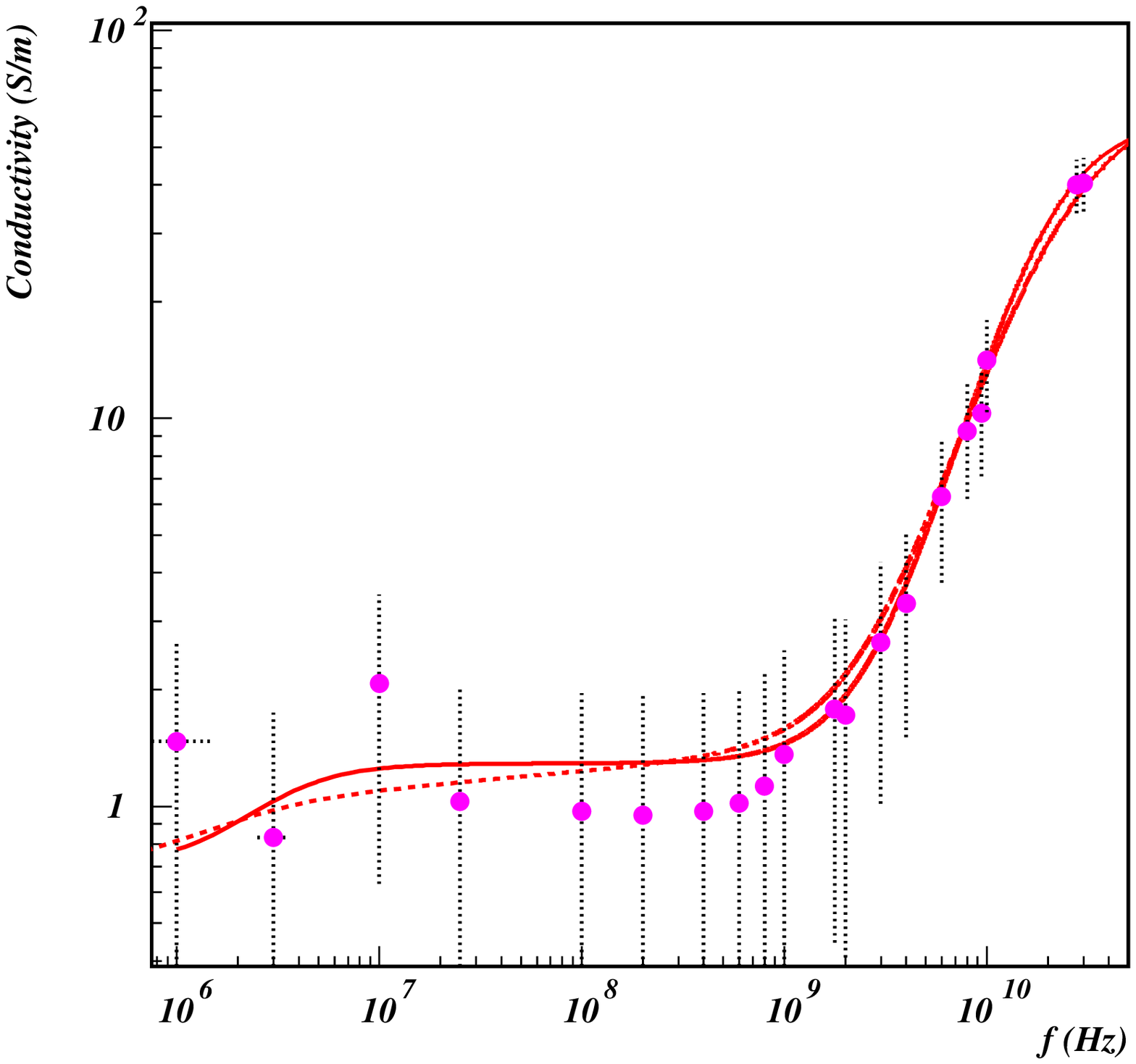}}
\vspace*{- 0.5cm}
\caption{\label{DCC} Relative permittivity (left) and conductivity (right) of 
blood parametrized by the Debye model (Equation~\ref{Dpar}; solid line) 
and the Cole-Cole model (Equation~\ref{CC}; dashed line).  
Data are from References \cite{Schw85,Hahn80,Cook52,Pfut52,Burd80,Alis93,Schw63}.}
\end{center}
\end{figure}

The Debye parametrization thus solves all the indicated problems associated 
with Cole-Cole parametrization.  It remains to be determined, however, whether 
the Debye approach also provides a sufficiently accurate description of 
physical properties of a biological material.  To ascertain this, we compared 
the Debye and Cole-Cole models in the case of blood.  As shown in Figure~\ref{DCC}, 
the two parameterizations describe equally well data from 
References \cite{Schw85,Hahn80,Cook52,Pfut52,Burd80,Alis93,Schw63}
in the frequency range 1 $MHz$-100 $GHz$, important for nanopulse research.  
It can be concluded that replacing the Cole-Cole model with the Debye model 
does not compromise the level of description of physical properties
of the material.

\section{Field Calculation}

Above we outlined an approach to applying FDTD to calculate an EM field 
based on the Debye model and compared it to the Cole-Cole model.  
Requirements include a description of the source field and of the geometry 
and electromagnetic properties of the material that is both accurate and 
suitable for computational modeling.  Now we present some results of 
calculations more specifically pertinent to nanopulse bioeffects research.

The cuvette shown in Figure \ref{fig05} was exposed to the EM pulse described 
by Equation~\ref{dubexp}. 
Electrical properties of the material inside the cuvette were described by 
Equation~\ref{Dpar}, explicitly written as

\begin{equation}
\epsilon(\omega) = \epsilon_{\infty} 
+{{\epsilon_{s1}-\epsilon_{\infty}} \over {1+i\omega\tau_{1}}}
+{{\epsilon_{s2}-\epsilon_{\infty}} \over {1+i\omega\tau_{2}}}.
\label{Dpar2}
\end{equation}
Parameters of materials used in the calculations are presented in Table ~\ref{tab1}.
The choice of materials was intended to provide a close approximation of 
the materials in the experimental work of the LA Tech-led collaboration.

\begin{table}
\begin{center}
\begin{tabular}{lcccccc}
\multicolumn{1}{c}{Material} {\vline}&
\multicolumn{1}{c}{$\epsilon_{\infty}$} {\vline}&
\multicolumn{1}{c}{$\epsilon_{s1}$} {\vline}&
\multicolumn{1}{c}{$\epsilon_{s2}$} {\vline}&
\multicolumn{1}{c}{$\tau_{1} (s)$} {\vline}&
\multicolumn{1}{c}{$\tau_{2} (s)$} {\vline}&
\multicolumn{1}{c}{$\sigma (S/m)$} {\vline}\\
\hline\hline
   Plastic            &  2.0  &  -     &  -   &  - & - &0.\\
   Water              &  4.9  &  80.1  &  -   & $10.0 \; 10^{-12}$ & - &0.\\
   Ionized Water      &  4.9  &  80.1  &  -   & $10.0 \; 10^{-12}$ & - & Variable\\
   Blood              &  7.0  & 4007.0 & 62.0 &  $6.0 \; 10^{-8}$  & $8.37 \; 10^{-12}$   &0.7\\
   Bone (Cancellous)  &  2.5  &  97.5  & 11.0 &  $1.5 \; 10^{-8}$  & $8.37 \; 10^{-12}$   &0.07\\
   Bone (Cortical)    &  2.5  &  37.5  &  5.5 &  $1.5 \; 10^{-8}$  & $8.37 \; 10^{-12}$  &0.02\\
\end{tabular}
\end{center}
\caption{Debye parameters for the materials used in the computation. Parameters for
water are based on Reference ~\cite{Chang79}.  Parameters for blood and bones are 
from a fit to data in Reference ~\cite{Gab96}.  Static conductivity, $\sigma$, 
is also from Reference ~\cite{Gab96}.}
\label{tab1}
\end{table}

\begin{figure}
\begin{center}
\mbox{\epsfxsize=3.in\epsffile{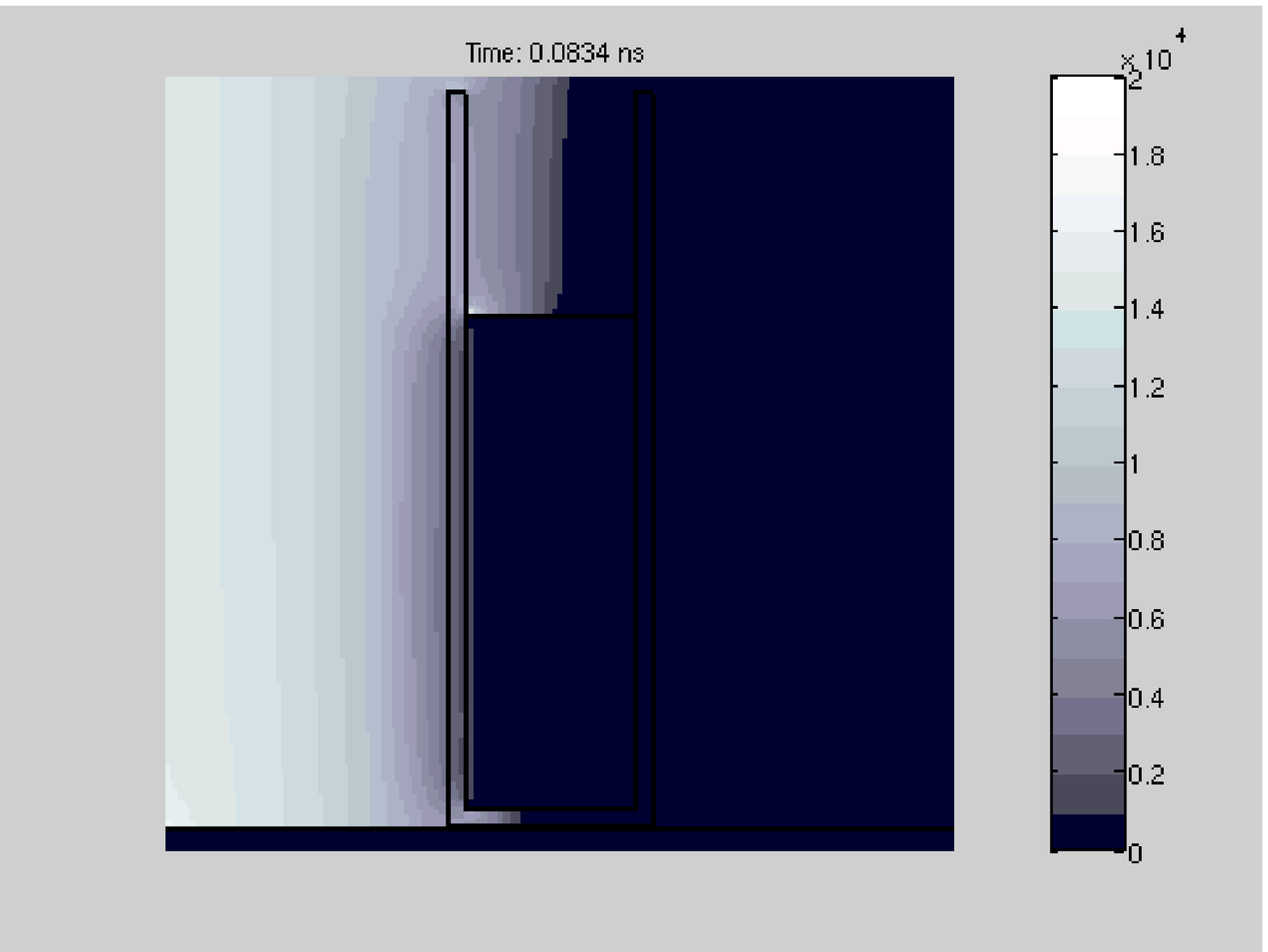}
\epsfxsize=3.in\epsffile{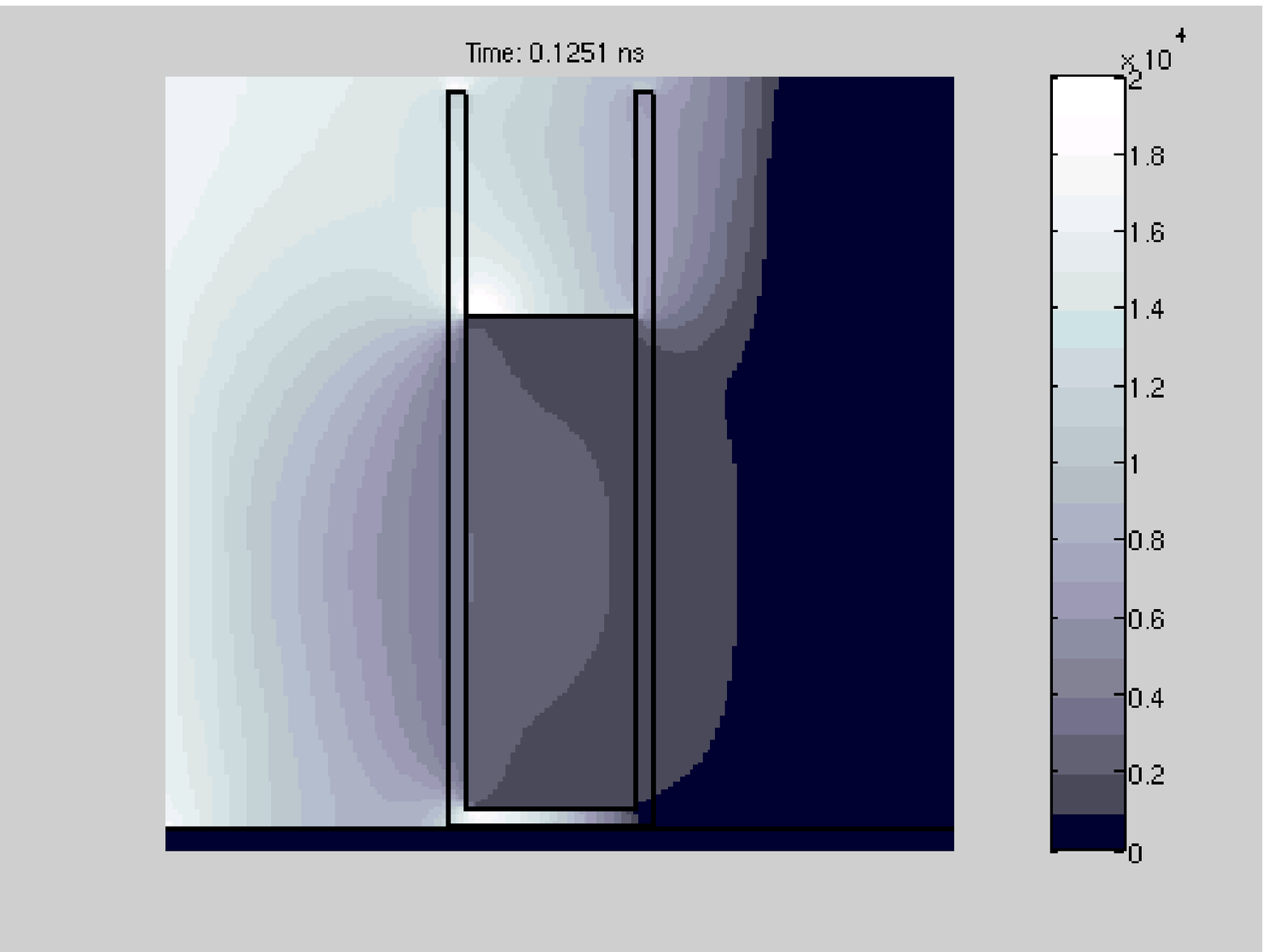}}
\mbox{\vspace*{-5.8cm}\epsfxsize=3.in\epsffile{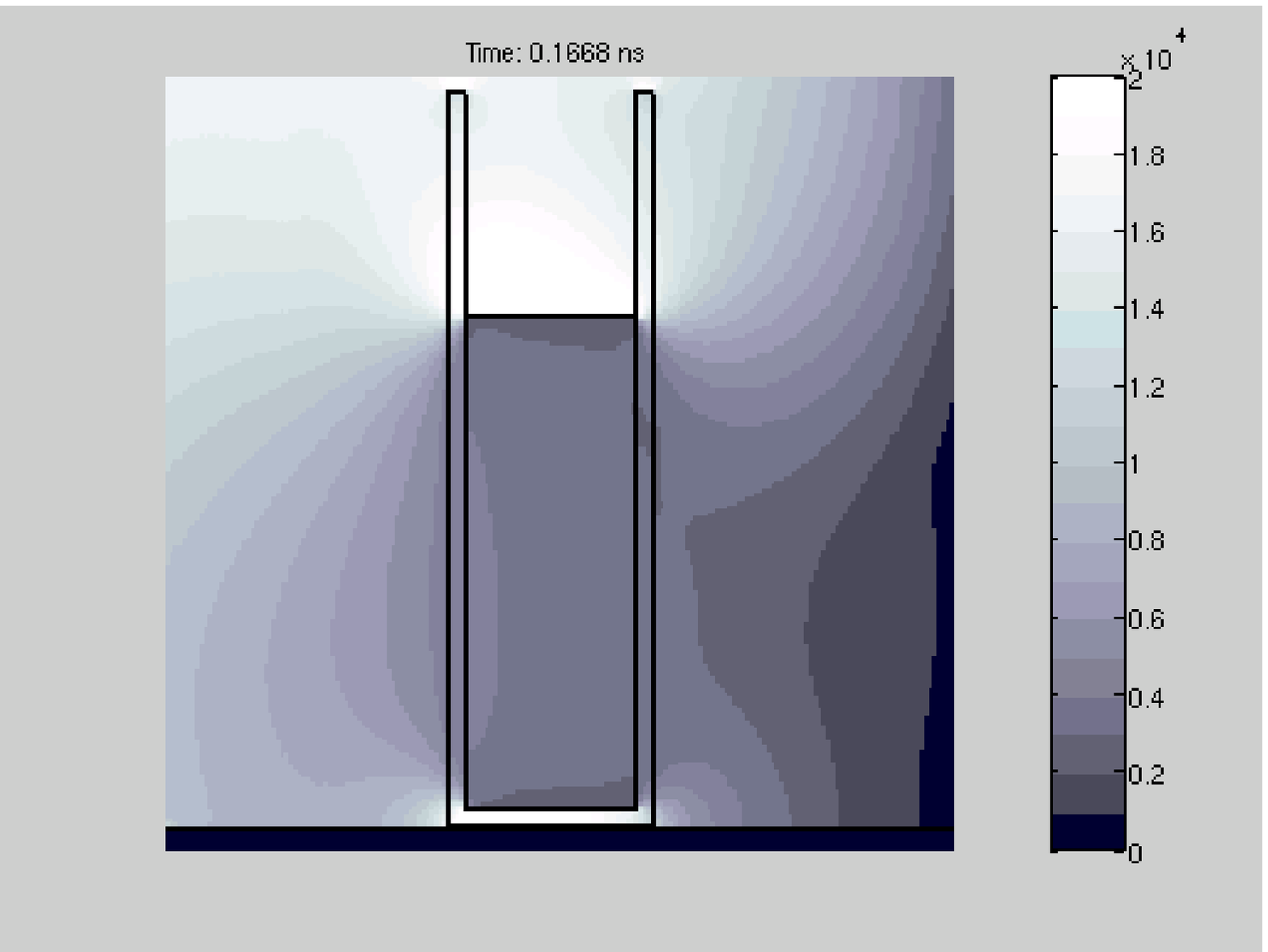}
\epsfxsize=3.in\epsffile{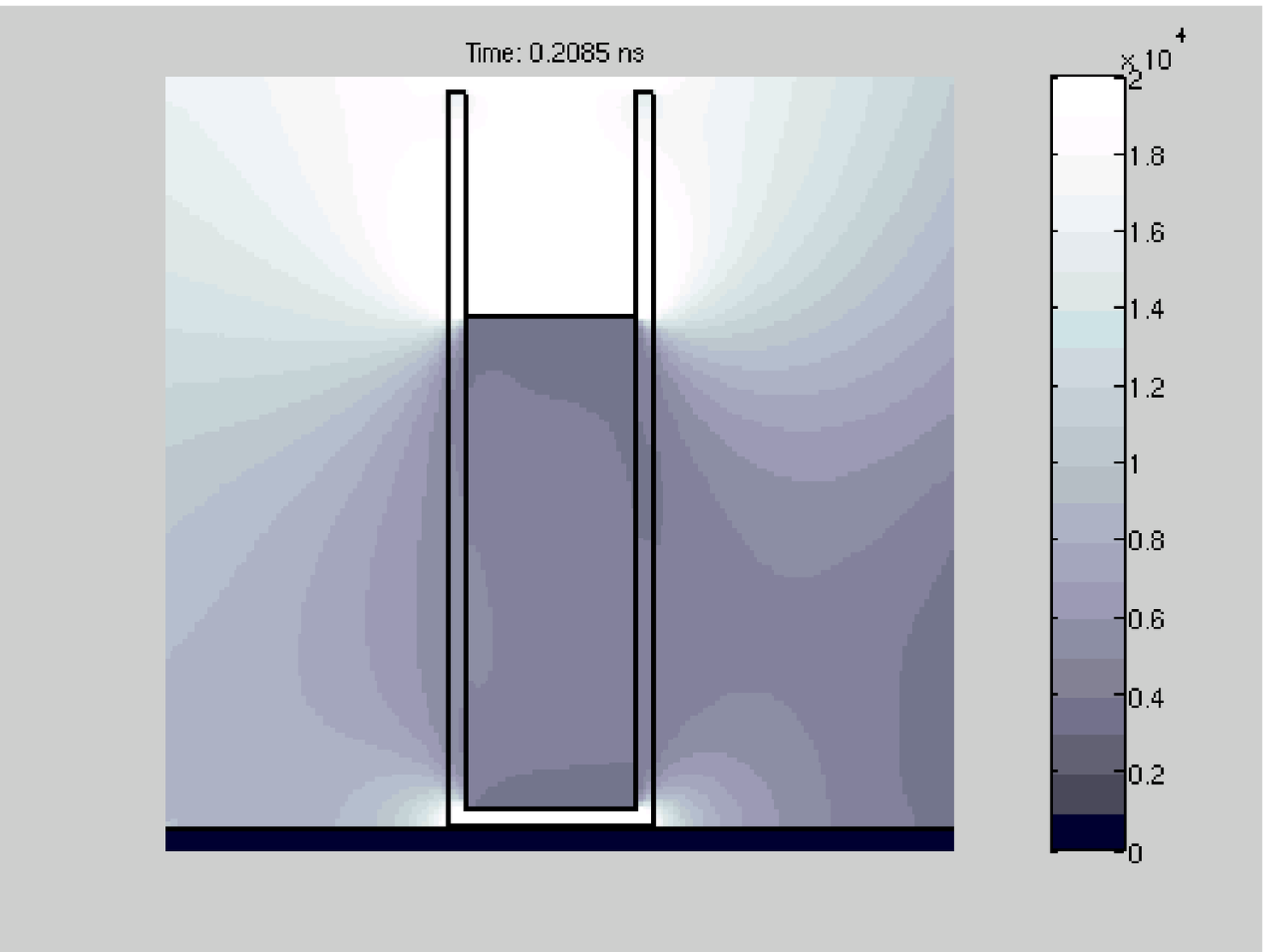}}
\end{center}
\caption{Penetration of an EM pulse into a blood-filled polystyrene cuvette.  
Contours represent the $y$-component of the electric field in steps of 1000 $V/m$.}
\label{snaps}
\end{figure}

FDTD calculations of exposure of a biomaterial to a nanopulse provide a description 
of the field throughout the time range.  This enables the creation of animated 
movies and analysis of the behavior of the EM field in time.  
Snapshots only can be presented here.  
As an example, Figure \ref{snaps} shows penetration of an EM pulse in a 
cuvette filled with blood.  The complete animation can be accessed 
on-line \cite{Neven04}.

\begin{figure}
\begin{center}
\mbox{\epsfxsize=3.0in\epsfbox{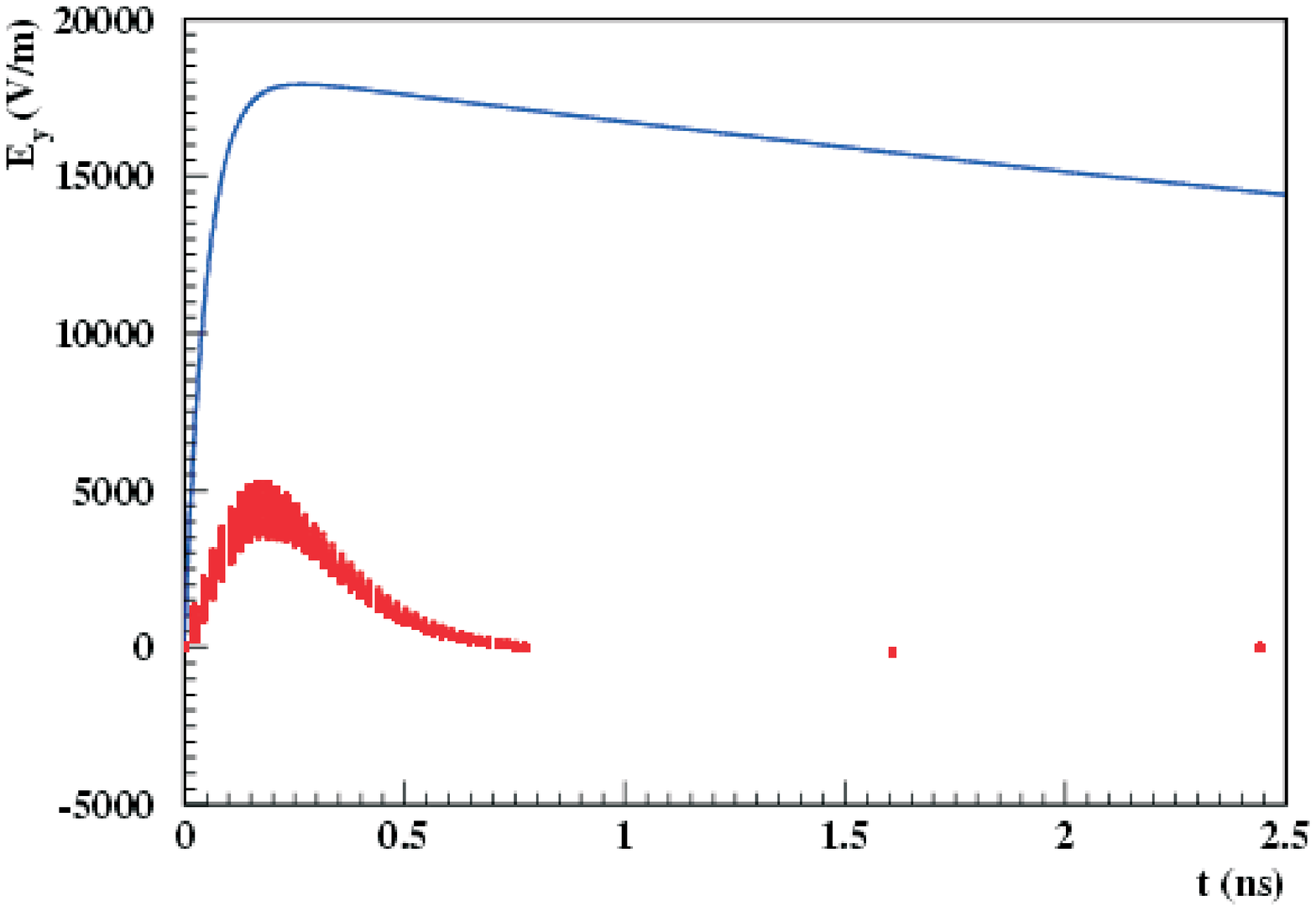}
\epsfxsize=3.0in\epsfbox{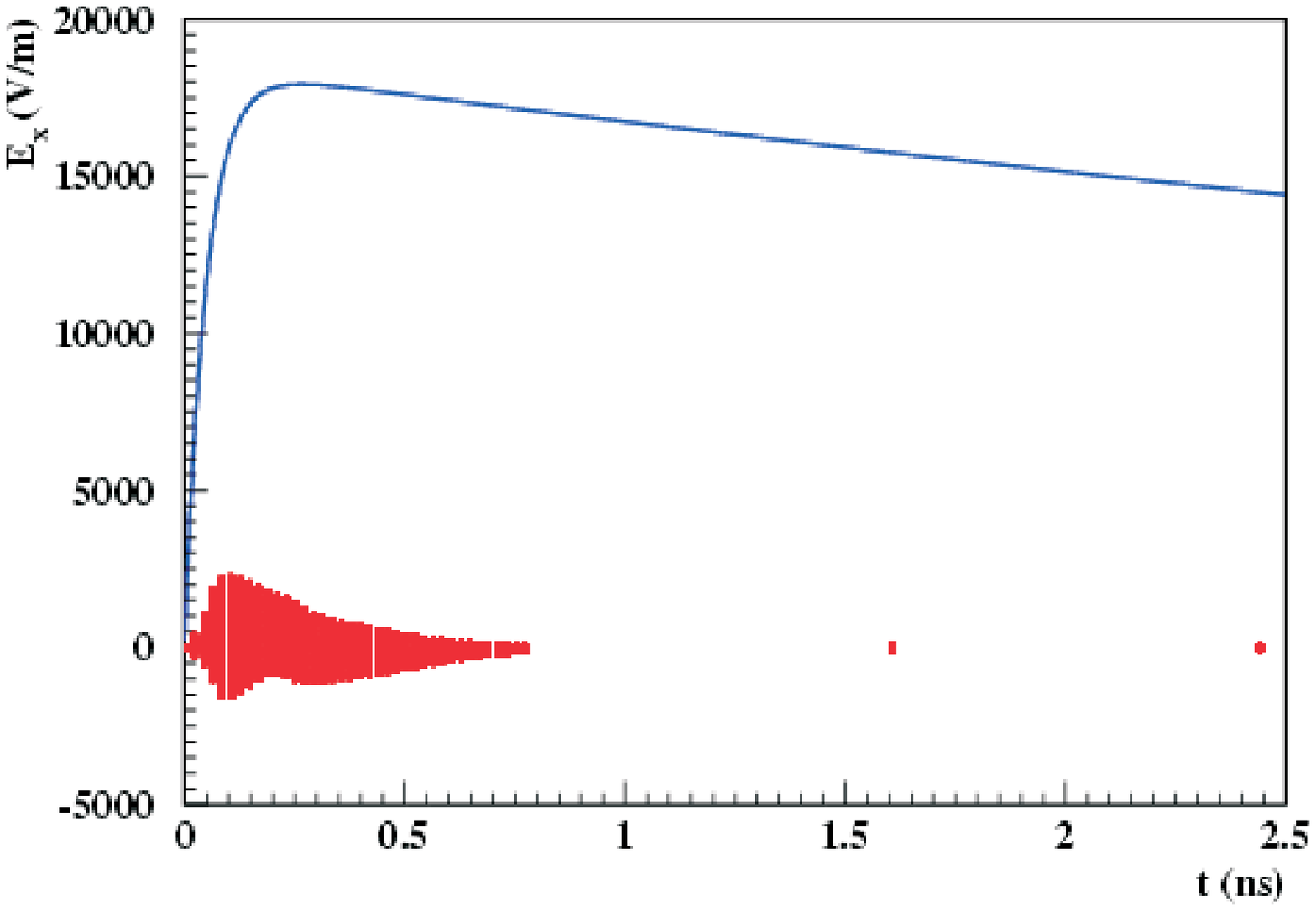}}
\vspace*{- 0.5cm}
\end{center}
\caption{Comparison of components of the electric field in the blood-filled 
cuvette to shape of the incident pulse for a span of 2.5 $ns$.  
Distribution of the field values in a particular time is a measure 
of the inhomogeneity of the field across the sample.}
\label{eblood}
\end{figure}

\begin{figure}
\begin{center}
\mbox{\epsfxsize=3.0in\epsffile{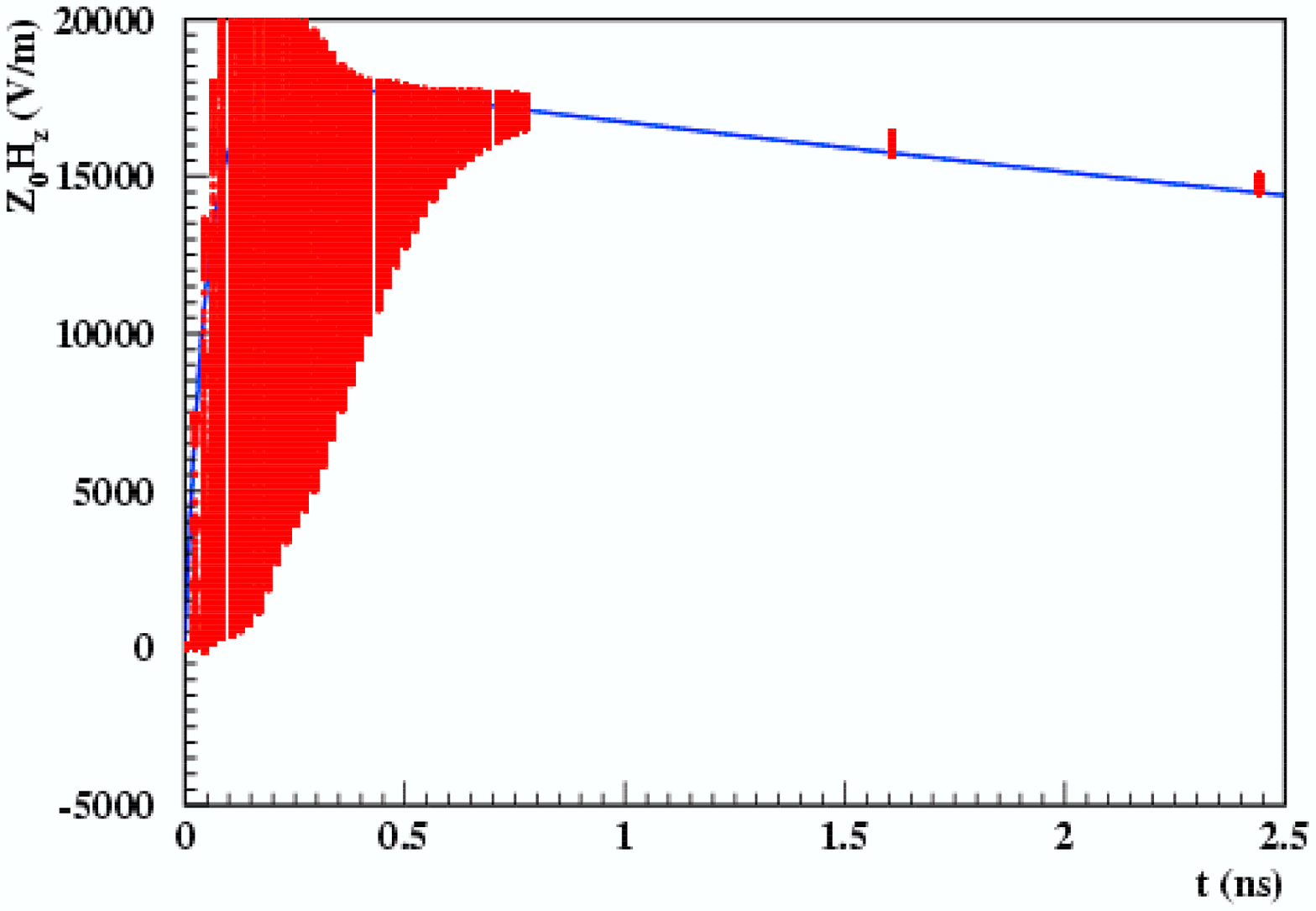}
\epsfxsize=3.0in\epsffile{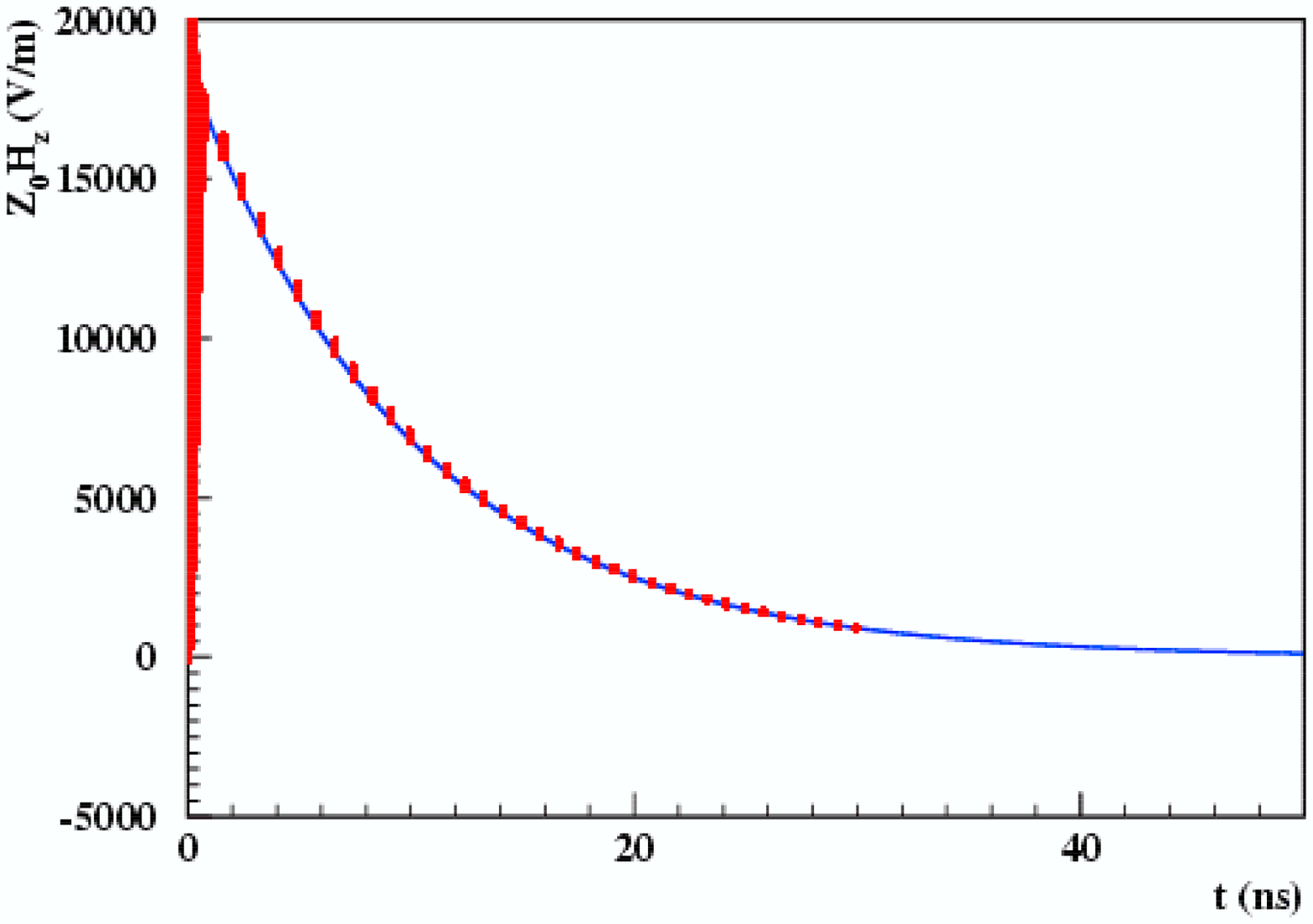}}
\vspace*{- 1.5cm}
\end{center}
\caption{Magnetic field component multiplied by the impedance of free space, 
$Z_{0}=376.7 \; \Omega$.  Left, field components (red vertical lines) are 
superimposed to the shape of the incident pulse for the first 2.5 $ns$.  Right, calculated data for 50 $ns$. The distribution of the field values in a particular time is a result of inhomogeneity of the field across the sample.}
\label{hblood}
\end{figure}

Properties of exposing the blood-filled cuvette to a linearly-polarized EM pulse 
described by Equation~\ref{dubexp} can be summarized as follows:

\begin{itemize}

\item Penetration of the electric component is defined substantially more 
by pulse rise time than pulse width, and the width inside the blood sample 
is an order of magnitude shorter than the width of the incident pulse 
(Figure \ref{eblood}).  The component of the electric field in the direction 
of polarization ($y$) is at least a factor of two larger than the component 
induced in the perpendicular direction ($x$).

\item The magnetic field component in the material is dominated at first 
by rise-time induction and then, as the penetrated electric field components 
fall to zero, behaves as though no material were present (Figure \ref{hblood}).

\end{itemize}

Ionized water of the conductivity of blood gave essentially the same result 
as blood. This means that in nanopulse research the dielectric properties of 
biological matter are dominated by those of water at high frequencies.  
It follows that model parameterization at high frequencies is important 
for describing the propagation of a nanopulse in biological matter.

\begin{figure}
\begin{center}
\mbox{\epsfxsize=3.in\epsffile{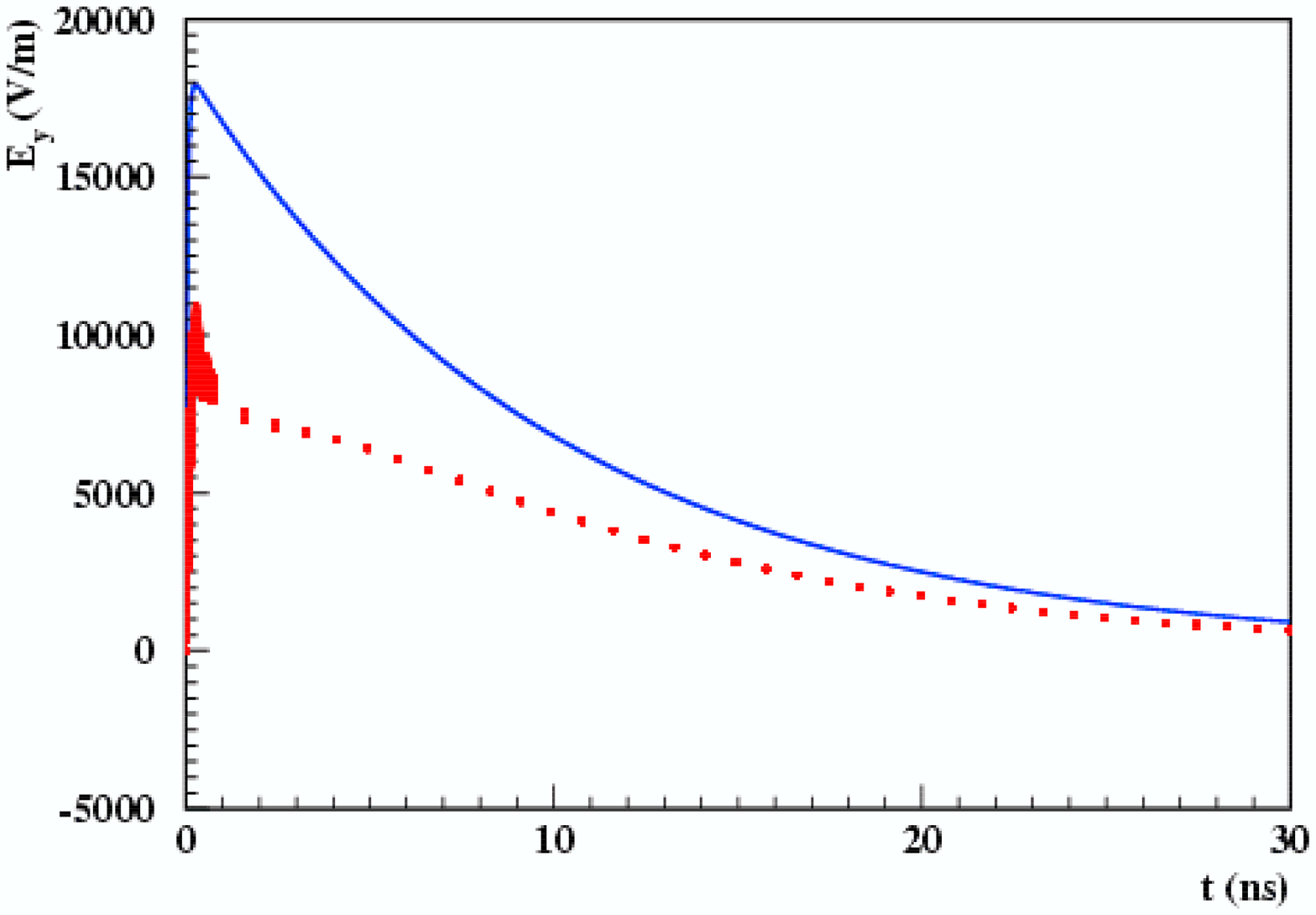}
\epsfxsize=3.in\epsffile{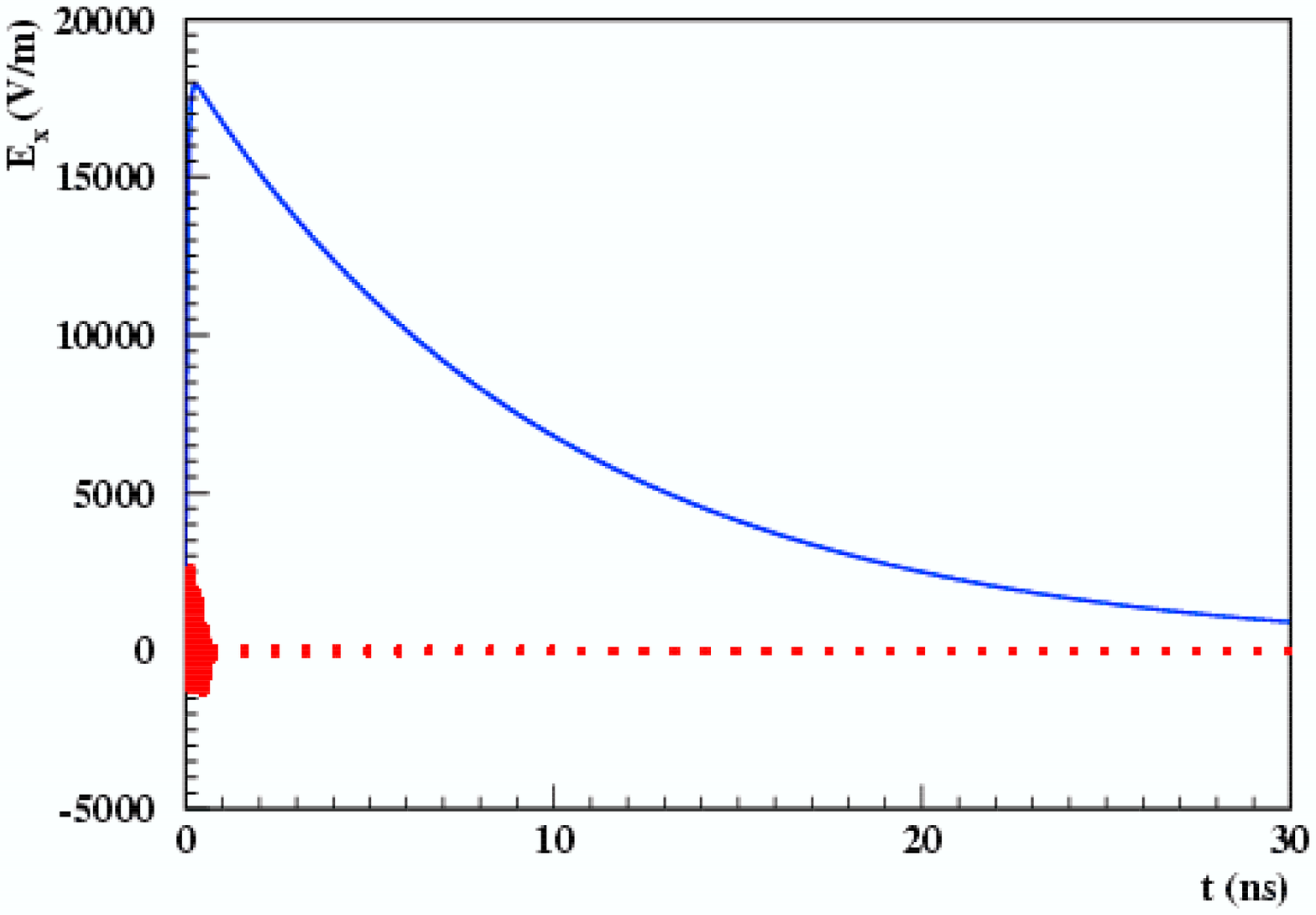}}
\vspace*{- 1.5cm}
\end{center}
\caption{Components of the electric field inside water (red) plotted with 
incident pulse (blue) for the first 30 $ns$.  The distribution of field components 
for a particular time point reflects the inhomogeneity of the field across 
the sample.}
\label{ewater}
\end{figure}

For pure water the situation can be summarized as follows:

\begin{itemize}

\item Penetration of the electric component in the direction of polarization 
($y$) is defined by both rise time and pulse width.  The pulse inside water 
is a superposition of a short pulse, induced by a fast rise time, and the 
longer incident pulse (Figure \ref{ewater}).

\item The electric field perpendicular to the direction of polarization ($x$) 
is defined by rise time only (Figure \ref{ewater}).

\item The magnetic field component is at first dominated by electrical induction, 
and, as the penetrated electric field components fall to zero, 
behaves as though no material were present, as in the case of blood.

\end{itemize}

\begin{figure}
\begin{center}
\mbox{\epsfxsize=3.in\epsffile{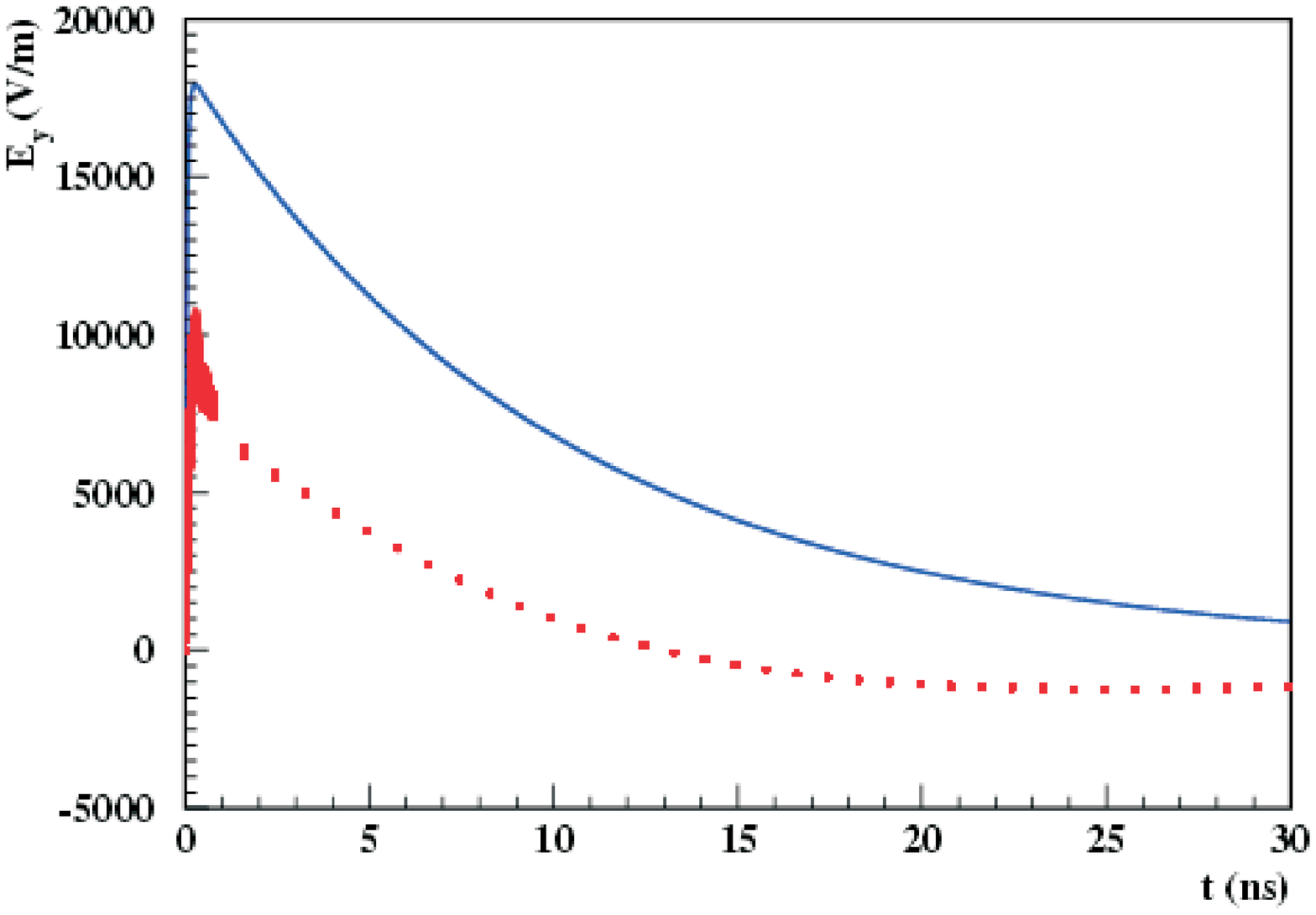}
\epsfxsize=3.in\epsffile{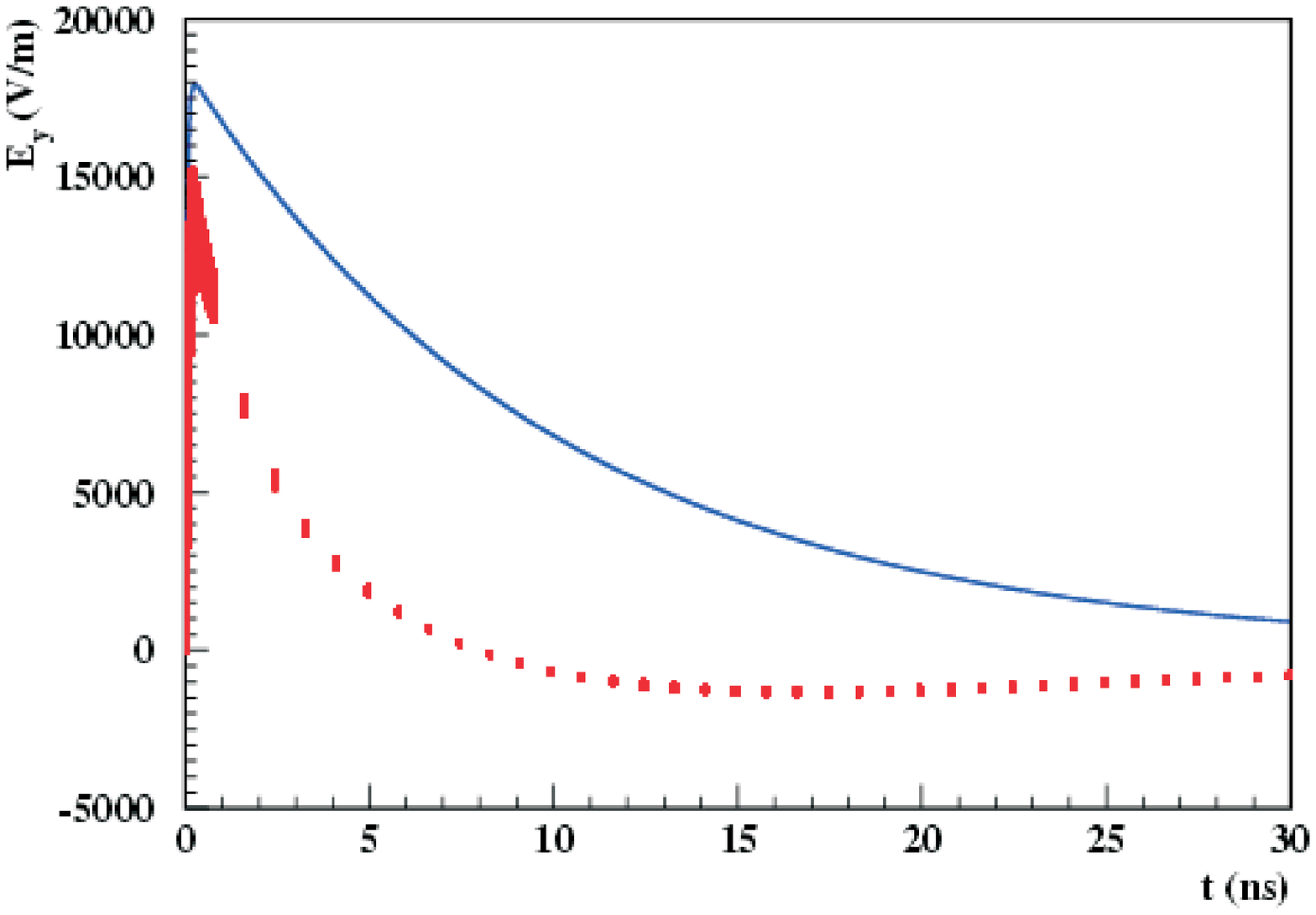}}
\vspace*{- 0.5cm}
\end{center}
\caption{Electric field in the direction of polarization in bacterial growth 
medium (left) and in cortical bone (right), plotted with the shape of 
the incident pulse in the first 30 $ns$. The distribution of the field values 
for a particular time point measures inhomogeneity of the field across the sample.}
\label{elb}
\end{figure}

Bacterial growth medium was simulated as water with a conductivity of 11.6 $mS/m$.  
The results agree with expectations based on the calculations on blood and water.  
The shape of the electric component in the direction of polarization is in 
essence similar to that for pure water.  The width, however, is shortened by 
the low conductivity, as shown in top panel of Figure \ref{elb}.  
The bottom panel shows the result of the calculation 
for cortical bone, the biomaterial least similar to water.

\begin{figure}
\begin{center}
\mbox{\epsfxsize=3.in\epsffile{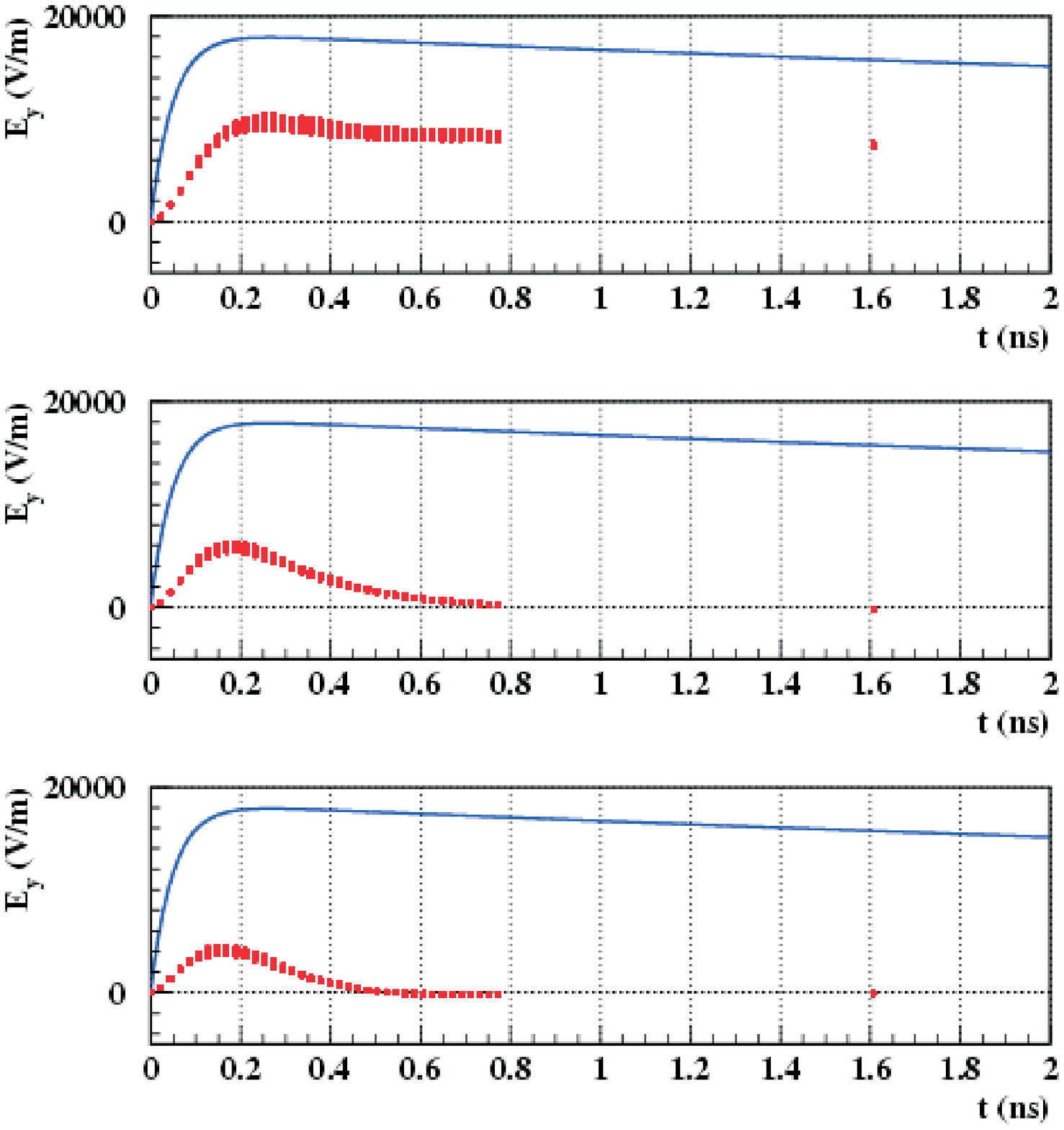}
\epsfxsize=3.in\epsffile{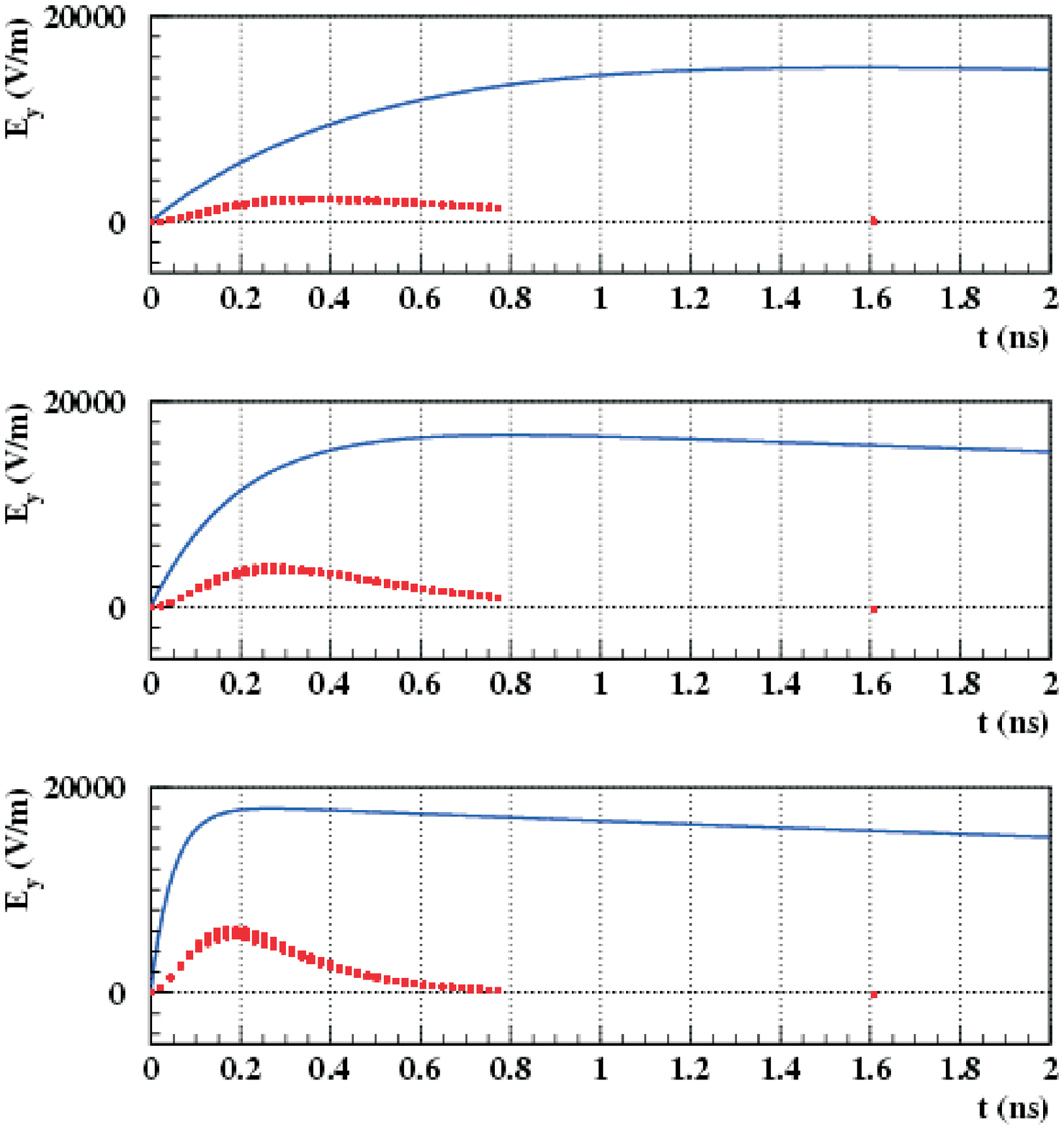}}
\vspace*{- .5cm}
\end{center}
\caption{Left side: Electric field in the direction of polarization inside water with a 
conductivity of 0 $S/m$ (top), 0.5 $S/m$ (middle), and 1 $S/m$ (bottom).  
The rise time of the incident pulse was 100 $ps$ in each case. Right side:
Electric field in the direction of polarization inside water of 
conductivity 0.5 S/m. The incident pulse rise time was 780 $ps$ (top), 
380 $ps$ (middle), and 100 $ps$ (bottom). }
\label{3cond}
\end{figure}

It became apparent in the course of this work that pulse penetration 
is a function of both rise time and pulse width. 
For a non-conductive material, both pulse features are important.  
For a conductive material, depending on conductivity, penetration is dominated 
by rise time. For blood, a material of considerable conductivity, incident 
pulse width is relatively unimportant.  Left side of Figure \ref{3cond} shows the 
penetration of a nanopulse inside a material as a function of conductivity.  
As conductivity increases amplitude and width of the penetrating pulse decrease; 
the pulse becomes a function of rise time only. In the right side of 
Figure \ref{3cond}, the conductivity of water was a constant 0.5 $S/m$ 
while the pulse rise time varied from 780 $ps$ to 100 $ps$.

FDTD also allows quick calculation of the pulse energy deposited in a 
biological material. Conversion of electromagnetic energy into mechanical or 
thermal energy is computed using \cite{Jackson99}

\begin{equation}
P= \int_{V} \vec J \cdot \vec E  \;dV,
  \label{power}
\end{equation}
where $P$ is deposited energy in unit of time, and $\vec J$ and $\vec E$ are, 
respectively, current density and electric field inside the material. 
FDTD provides the values of $\vec E$ and $\vec J$ (from Equation \ref{JE}) 
through the entire volume at any time. Numerical integration of Equation \ref{power},
used to determine the amount of energy deposited per pulse, is straightforward.  
The results show that this energy is small and does 
not influence the temperature of the exposed material for the pulse repetition 
rates of the order of few $MHz$ or less. The average converted energy per pulse 
of the pulse described by Equation~\ref{dubexp} was $\sim 0.003 \; J/m^{3}$ for 
blood and $\sim 0.0005 \; J/m^{3}$ for water.  The resulting temperature increase, 
about $\sim 10^{-10} \; K$ per pulse, is clearly negligible.

\begin{figure}
\begin{center}
\vspace*{- 1.cm}
\mbox{\epsfxsize=3.0in\epsffile{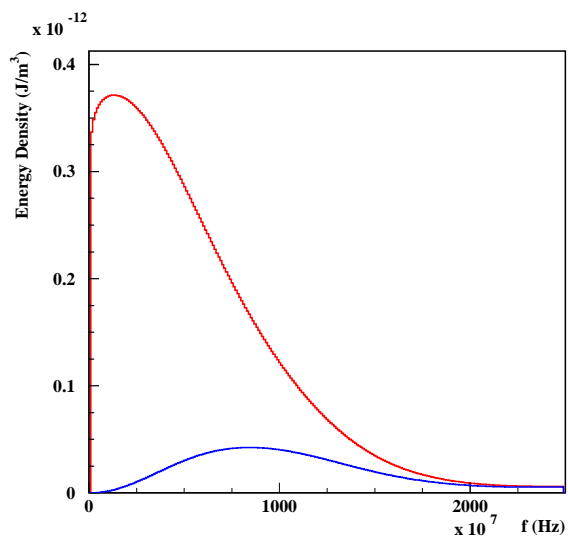}}
\end{center}
\vspace*{- 2.5cm}
\caption{Spectral energy density for blood (upper red curve) and water 
(lower blue curve). 
The integral of a distribution over all frequencies is the total pulse energy.}
\label{eden}
\end{figure}

Finally, the power spectrum or spectral energy density must be modeled to 
understand the interaction of short EM pulses with biological material.  
The spectrum for the cases of blood and water, obtained by Fourier transformation 
of Equation \ref{power}, is shown in Figure~\ref{eden}.

\section{Conclusion}

We have presented a series of results of FDTD calculations on nanopulse 
(ultra-wideband) penetration of biological matter.  Calculations included 
a detailed geometrical description of the material exposed to nanopulses, 
which is typically contained inside a cuvette or a Petri dish in 
an exposure chamber ($e.g.$ GTEM cell), and a state-of-the-art description 
of the physical properties of the material.  To ensure that the results 
would be sound, the length of a side of the Yee cell was set at 1/4 $mm$, 
smaller than the value required by the cut-off frequency of 100 $GHz$, 
and the Cole-Cole parametrization of the dielectric properties of tissue 
in the frequency range $\leq 100 \; GHz$ was used to describe the exposed material.
To minimize computation time, the Cole-Cole parametrization was reformulated in 
terms of the Debye parametrization with no loss of accuracy of description.  
In 2-dimensional FDTD, the decreased computation time enabled comparison of 
different materials on exposure to nanopulses. The results can be summarized 
as follows:

a) The shape of a nanopulse inside a biomaterial is a function of both rise time 
and width of the incident pulse. The importance of the rise time increases 
and becomes dominant as the conductivity of the material increases.

b) Biological cells inside a conductive material are exposed to pulses defined 
by rise time only, which is often substantially shorter than the duration of 
the incident pulse.  It is possible to define the pulse inside the material 
by the conductivity of the material and the rise time of the incident pulse.

c) The amount of energy deposited by the pulse is so small that no effect 
observed on exposure of a biological sample to nanopulses of $ \sim 20 \; kV/m$ amplitude will have a 
thermal origin.

Calculation of the electric field surrounding a biological cell is the first step 
in understanding any effect resulting from exposure to nanopulses.  
Fast and accurate numerical programs are necessary not only for such computation 
but also for optimization of future experiments.  Results of the 2-dimensional FDTD calculations reported here have been compared in selected cases with the 
full 3-dimensional calculation.  No significant difference in pulse 
propagation has been found thus far.  Graphical results of the full 
3-dimensional computation will be reported in a subsequent paper.

\section*{Acknowledgments}

We thank Weizhong Dai, Shengjun Su, and other members of the research team for helpfull discussions.
 
This material is based on research sponsored by the Air Force Research Laboratory, 
under agreement number F49620-02-1-0136. The U.S. Government is authorized to 
reproduce and distribute reprints for Governmental purposes notwithstanding any 
copyright notation thereon. The views and conclusions contained herein are those 
of the authors and should not be interpreted as necessarily representing 
the official policies or endorsements, either expressed or implied, of the 
Air Force Research Laboratory or the U.S. Government.

\end{document}